# The cosmic-ray ground-level enhancements of 29 September 1989 and 20 January 2005

**H. Moraal[1]**

*Centre for Space Research, School for Physical and Chemical Sciences, North-West University, Potchefstroom, 2520, South Africa.*

*E-mail:* harm.moraal@nwu.ac.za

**R.A. Caballero-Lopez**

*Ciencias Espaciales, Instituto de Geofisica, Universidad Nacional Autónoma de México, 04510 México D.F., México.*

*E-mail:* rogelioc@gmail.com

**K.G. McCracken**

*Institute for Physical Science and Technology, University of Maryland, College Park, MD, 20742, U.S.A*

*E-mail:* jellore@hinet.net.au

Enhancements of the comic-ray intensity as observed by detectors on the ground have been observed 71 times since 1942. They are due to solar energetic particles accelerated in the regions of solar flares deep in the corona, or in the shock front of coronal mass ejections (CMEs) in the solar wind. The latter is the favoured model for the classical "gradual" ground level enhancement (GLE). In several papers since the one of McCracken et al. (2008), we pointed out, however, that some GLEs are too impulsive to be accelerated in the CME shocks. This hypothesis, together with other properties of GLEs, is demonstrated graphically in this paper by plotting and comparing the time profiles of GLEs 42 of 29 September 1989 and GLE 69 of 20 January. These two events are respectively the largest examples of gradual and prompt events.

---

[1]Speaker

1. **Introduction.**

Ground-level enhancements (GLEs) in the intensity of cosmic rays as measured by neutron monitors are associated with solar flares and coronal mass ejections (CMEs). They originate primarily from western longitudes on the surface of the sun. Miroshnichenko et al. (2013) gave a recent review of these events.

In an accompanying paper (Moraal et al., 2015a), we study the time structure of these events. They range from "impulsive" or "prompt", with rise times as short as 5 minutes, to "gradual", with rise times up to 150 minutes. The hypothesis in that paper (and references therein) is that flares in the low solar corona are relatively short-lived and have dimensions much smaller than one solar radius. On the other hand, CMEs develop more gradually at distances beyond about four solar radii, are much larger than the sun; they should therefore have shock fronts that are widely extended in heliolatitude and longitude. It is therefore natural to associate the impulsive events with acceleration in solar flares, and gradual events in CME shock fronts.

In this paper we select the second and third largest GLEs, those of 29 September 1989 (GLE 42) and 20 January 2005 (GLE 69) to give a qualitative and graphical representation of their properties, which were studied quantitatively in the series of papers referenced in Moraal et al. (2015b). The two events are shown in Figure 1. GLE 42 is the best-observed example of a gradual event, and GLE 69 of a prompt event.

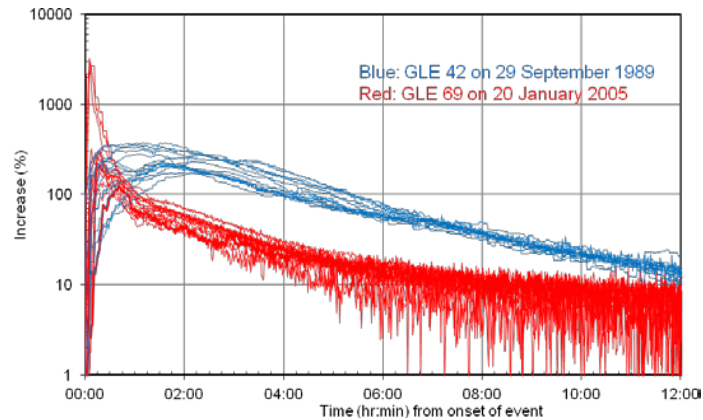

**Figure 1.** Ground-level enhancements on 29 September 1989 (blue) and 20 January 2005 (red). For GLE 42 the following stations are used: Apatity, Calgary, Cape Schmidt, Deep River, Goose Bay, Inuvik, Mawson, McMurdo, Mirny, Ottawa, Sanae (corrected to sea level), Terre Adelie, Thule, Tixie Bay. For GLE 69 the stations are: Apatity, Barentzburg, Cape Schmidt, Fort Smith, Inuvik, Mawson, McMurdo, Nain South Pole (corrected to sea level), Oulu, Terre Adelie, Tixie Bay, Thule. All these stations have geomagnetic cutoff rigidities < 1 GV.

Only the increases for stations with geomagnetic cutoff rigidity $P_c$< 1 GV are plotted in Figure 1. This eliminates energy or rigidity dependence in the comparisons, because the lowest observable rigidity for all stations is the atmospheric cutoff of ≈ 1 GV. The differences seen by the neutron monitors are then solely due to different directions of viewing, which is a tool to study the anisotropy of the event - the neutron monitor that sees the largest increase is aligned nearest to the direction of propagation of the anisotropic beam.

Moraal and Caballero-Lopez (2014) interpreted the combined profiles of GLEs 42 and 69 as that in both of them there was an initial, prompt acceleration, lasting only a few minutes. Thereafter there was a more gradual acceleration lasting several tens of minutes to more than an hour. In GLE 42 the effect of the prompt, initial acceleration was weak because the disturbance on the sun and its assumed associated flare were hidden behind the western limb. However, the ensuing CME had a very large angular extent, so that the particles accelerated in this second boost were readily detectable at Earth. In the case of GLE 69, the flare site at ~60° W was almost perfectly connected to Earth via the Parker spiral magnetic field, so that the promptly accelerated particles were very well visible at Earth, while those that were accelerated by the secondary mechanism remained engulfed below these particles until about an hour into the event.

## 2. Magnitude of the events

These two events are the second and third largest of the 67 events observed by neutron monitors since 1956. Four more GLEs were observed by ionisation chambers prior to this; two in 1942, and one each in 1946 and 1949. It is generally accepted that GLE 05 on 23 February 1956 was the largest one ever observed.

The strength of an event is usually characterised by the amplitude, or maximum increase observed. According to this measure, Figure 1 shows that GLE 69 is 12 to 15 times stronger than GLE 42. This measure was used by Bieber et al. (2013) and Miroshnichenko et al. (2013), who found that the relative strength of the events as observed by neutron monitors was in the ratio GLE 05: GLE 69: GLE 42 ≈ 1.0; 1.0; 0.08. (It must, however, be borne in mind that GLE 05 was observed by the Leeds neutron monitor at cutoff rigidity $P_c = 2.2$ GV, and would have been larger if it were observed by polar neutron monitors.)

The total number of particles with rigidity greater than the cutoff rigidity that reach a neutron monitor sees during an event is called the fluence. We argue that this is a more appropriate measure of strength, because it is an indicator of the total amount of energy imparted to charged particles during the event. This fluence is proportional to the area underneath each curve. For GLE 42 the highest curve in Figure 1 is for Thule, and for GLE 69 it is for Terre Adelie.

According to this fluence measure for the first 12 hours after onset, GLE 42 was 1.53 times larger than GLE 69.

To put the magnitude of these two events in further perspective, the largest and fastest riser for GLE 05 on 23 February 1956 was the Leeds neutron monitor, with an equivalent fluence 2.94 times that of GLE 42. However, this station was at cutoff rigidity 2.2 GV. If the spectrum was of the form $P^{-\gamma}$ with $\gamma = 5\pm1$, the calculations of Caballero-Lopez and Moraal (2012) imply that the fluence at 1 GV would have been 4±1 times that of GLE 42.

Hence, according to the fluence measure, the three largest GLEs observed in the neutron monitor era are in the ratio GLE 05: GLE 42: GLE 69 = 1.00±0.25: 0.25: 0.16.

## 2. Anisotropy

Neutron monitors are directionally sensitive. The atmosphere directs maximum sensitivity towards the zenith direction, and the geomagnetosphere causes bending of the particle trajectories such that each neutron monitor has a unique "asymptotic cone of acceptance" for particles outside the magnetosphere. (see, e.g. the review of Smart et al., 2000). Stations with viewing directions that are best aligned with the direction of an anisotropic beam of particles see the largest increase.

Figure 2 shows the highest and lowest risers for each of the two events. The difference between these is an indicator of the anisotropy of the events, which diffes drastically. For GLE 69, for instance, the ratio of South Pole to Thule during the first ~10 minutes (not fully resolved on the plot) is so large that it is indeterminable. This indicates a highly collimated beam of particles, well-aligned with the asymptotic direction of viewing of the South Pole neutron monitor, and opposite to the viewing direction of Thule. For GLE 42 the difference between the highest and fastest riser (Thule) and the lowest and slowest riser (Sanae) during the first ~ 30 minutes is about 10 times smaller than for GLE 69. This indicates a much more angularly extended beam of particles. In Moraal and Caballero-Lopez (2014) we have used this difference in anisotropy as one of the inferences that GLE 42 was more likely accelerated in a gradual, spatially extended CME shock, while the peak of GLE 69 occurred so fast that it was more likely related to acceleration in a prompt, short-lived solar flare.

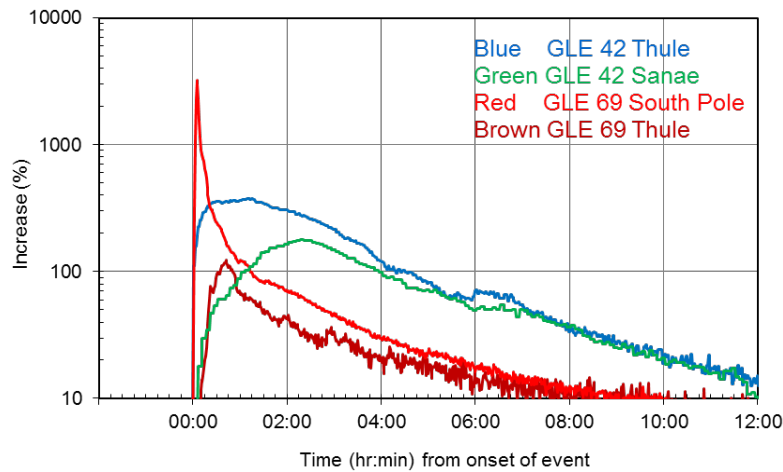

**Figure 2**. Fastest and slowest risers for the two events. Notice that about one hour into the event the slow-rising GLE 42 has a larger anisotropy than the fast-rising GLE 69.

We note exceptions to this general anisotropic behaviour. For a short period at 00:40 min. the ratio between Terre Adelie and Thule for GLE 69 is only a factor 1.5 times. This is much smaller than the Thule to Sanae ratio of 5.6 at the same time for GLE 42. This indicates the high temporal variability of the anisotropy. For both events the anisotropy diminishes gradually, to become insignificant after about 08:00 hours.

### 3. Multiple peaks

Figure 1 gives a clear indication of fluctuations in the time-intensity profiles. Such fluctuations can be due to multiple sources, multiple acceleration boosts, and multiple particle releases in the vicinity of the sun, but also due to fluctuations in the direction of the helisopheric magnetic field along which the particles propagate from the sun to Earth.

McCracken et al. (2008) interpreted the first peak at Sanae (reaching a maximum within the first 5 minutes) in GLE 69 as due to flare acceleration, but the remainder of the event as due to CME shock acceleration. The second and third peaks at 19 and 36 minutes after the onset were not attributed to multiple acceleration boosts or releases, but rather due to HMF fluctuations, which cause swings in the direction of beam propagation. By association with the first pulse at Sanae, the larger (but more delayed) first peak at Fort Smith was also interpreted as due to a flare source.

The increases of GLE 42 also show multiple peaks. For example, Inuvik (blue) has two peaks on both sides of a dip at ≈ 01:30, and McMurdo (green) two peaks on both sides of a dip at ≈ 02:00. Miroshnichenko et al. (2000) considered the Inuvik behaviour as "non-classic" and that of McMurdo as peculiar, because they identified it as separate peaks. However, they then gave a description of the directional sensitivity of neutron monitors, from which they concluded that such fluctuations can indeed be interpreted as due to fluctuations in HMF direction.

We point out that such multiple peaks – as indeed the entire profile of the GLE - are distorted out of proportion on linear plots, which is the custom for GLEs. This makes the McMurdo time profile look like two distinct peaks. The diffusive solution of Duggal (1979), (see also Moraal et al., 2015a,b#) indicates, however, that the increases should be exponential in time, while the decreases are somewhere between exponential and power-law form. The exponential plots we use therefore keep these fluctuations in much better relative perspective, and the GLE 42 fluctuations in Figure 3 does not give the impression of double peaks.

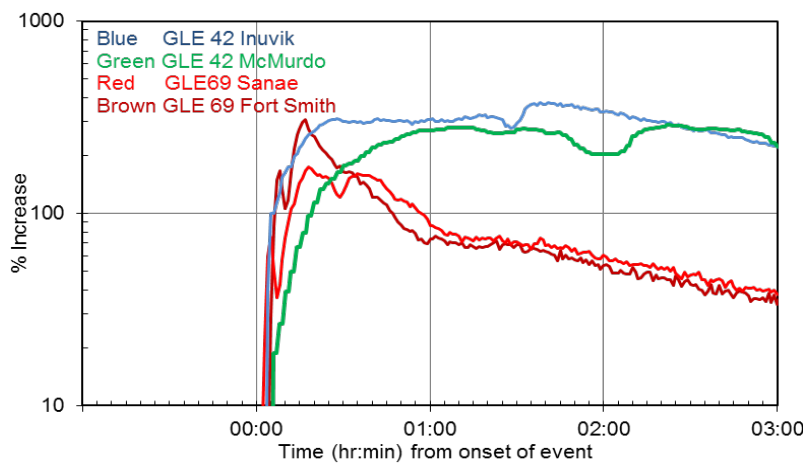

**Figure 3**. Multiple peaks. For GLE 69, Sanae (red) has three peaks, at ~ 5 min., 19 min. and 36 min. Fort Smith (brown) shows two. For GLE 42, Inuvik (blue) has two peaks on both sides of a dip at ≈ 01:30, and McMurdo (green) two peaks on both sides of a dip at ≈ 02:00. Only the peak at ~ 5 min. at Sanae for GLE 69, and possibly the peak at ~ 15 min at Fort Smith are interpreted as due to a first, prompt, and short-lived source. The other peaks are due to HMF fluctuations.

A second reason why the majority of these multiple peaks should not be due to multiple accelerations or releases is that they occur differently, almost randomly, on different neutron monitors. This can be seen on, especially the blue curves for GLE 42 in Figure 1. This behaviour is naturally explained as viewing directions that swing into and out of the beam. If they were due to multiple accelerations and/or releases, these fluctuations would occur in unison. More detail of the multiple peaks in GLE 42 are shown in Figure 15 of Moraal and Caballero-Lopez (2014).

This highlights the importance of looking at all the available increases together, instead of the usual selection of a few neutron monitors.

## 4. The use of bare counters alongside standard neutron monitors

Proportional neutron counters are only about 10% effective to register thermal neutrons, with the exact number dependent on the specific type and design. To boost the counting rate, the proportional counters in a neutron monitor are embedded in lead, which multiplies the number of incident neutrons by a factor of about 10 (also strongly dependent on design). Hence, the typical neutron monitor registers about one count per incident neutron. Design detail of the neutron monitor can be found in Hatton and Carmichael (1964).

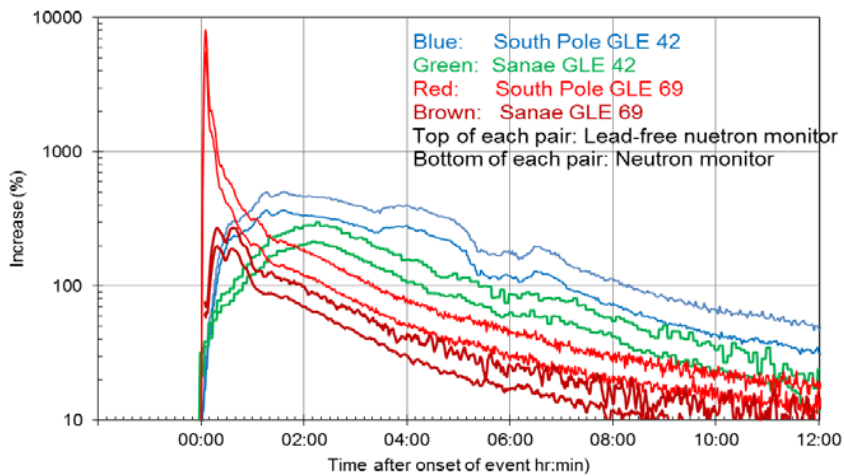

**Figure 4.** The events as seen by the neutron monitor (NM) and lead-free neutron monitor (LFNM) pairs at South Pole and at Sanae

The lead multiplier alters the energy sensitivity of the counters somewhat, so that it responds to slightly higher energies than a bare counter. This is embodied in different so-called yield functions for the two instruments (see, e.g., Caballero-Lopez and Moraal, 2012 and references therein).

The standard six-counter NM64 neutron monitor with $^{10}BF_3$ counters at Sanae has been accompanied by four lead-free counters of the same design since 1971. We refer to this as the lead-free neutron monitor or LFNM. Similarly, the three-counter NM64 neutron monitor at South Pole has had a six-counter $^{10}BF_3$ LFNM since 1989. This operated until the end of 2005, and it was re-commissioned in 2010 with 10 $^3$He counters.

Figure 4 shows the response of these LFNMs to GLEs 42 and 69, in comparison with the neutron monitors on the same site. Figure 5 shows that the typical LFNM/NM ratio is 1.5. Moraal and Caballero-Lopez (2014) used these ratios to determine the spectral index of GLE 42. McCracken et al. (2008) used this ratio as observed at Sanae to deduce that the first peak in the first 10 minutes of the event was harder than the remainder. Bieber et al. (2002) used this ratio to deduce the spectral index for GLE 59 on 14 July 2000.

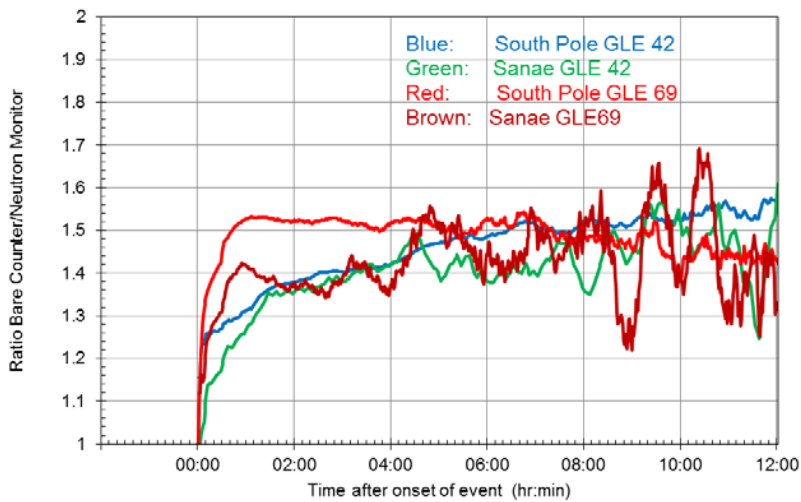

**Figure 5.** The LFNM/NM ratios at Sanae and at South Pole.

In principle such calculations can be done equally well with pairs neutron monitors with different cutoff rigidities, but there are two drawbacks. First, different environmental conditions (pressure, temperature, humidity, altitude) require corrections which introduce uncertainties. The NM and LFNM pair at a particular station, however, experience identical environmental conditions so that no corrections are needed. Second, neutron monitors at different locations have different asymptotic cones of acceptance. This implies that an anisotropy correction must be made before spectral calculations can be done, which introduces further uncertainties.

This anisotropy effect also applies to the NM-LFNM pair at the same location. However, most polar neutron monitors have narrow cones of acceptance, which means that the particles of all rigidities that produce the counts come from essentially the same direction. Therefore, the LFNM/NM ratio is a pure and clear signal of the energy/rigidity dependence of the event.

This argument does, however, not pertain to the Sanae neutron monitor (or other low-latitude neutron monitors). At this location the asymptotic cone of acceptance is wide. It is such that low-rigidity particles, just above the cutoff rigidity come from much farther East than the high–rigidity particles. This effect explains the difference in the LFNM/NM ratio for GLE 69 as observed by South Pole (red) and Sanae (brown). Caballero-Lopez and Moraal (2015) showed how the argument can then be changed around, namely that the difference between these two pairs becomes a measure of anisotropy – once again free of the uncertainties introduced by environmental effects.

In an accompanying paper at this meeting, Usoskin et al. (2015) used this concept to deploy two mini neutron monitors, of which one is a LFNM, at Dome C in Antarctica form the beginning of 2015. The design of these mini neutron monitors is described in Krüger et al. (2008), Krüger and Moraal (2010), and Krüger et al. (2015).

**Acknowledgements**

This work was supported by South African NRF Grant SNA2011110300007, and Mexican PAPIIT-UNAM grant IN110413. K.G. McCracken acknowledges the consistent support he has received since 2005 from the International Space Science Institute (ISSI), Bern, Switzerland. The neutron monitor observations used are from the worldwide network, collected in the database described in McCracken et al. (2012) http://usuarios.geofisica.unam.mx/GLE_Data_Base/files/.